\documentclass[aip, jcp, notitlepage, floatfix, reprint,
  citeautoscript, twocolumn]{revtex4-1}

\usepackage{amsmath}
\usepackage{amsfonts}
\usepackage{graphicx}
\usepackage{dcolumn}
\usepackage{miller}
\usepackage[table]{xcolor}
\usepackage[version=3]{mhchem}
\usepackage{transparent}
\usepackage{gensymb}
\usepackage{tabularx}
\usepackage{etoolbox}
\usepackage[hidelinks]{hyperref}
\usepackage{bm}
\usepackage{relsize}
\usepackage[normalem]{ulem}
\usepackage{cases}

\def\maxfloatwidth{%
  \ifdim\columnwidth>246.0pt
  300.0pt  \else
  \columnwidth
  \fi
}

\newcommand{\mbf}[1]{\mathbf{#1}}
\newcommand{\mrm}[1]{\mathrm{#1}}

\newcommand{\tcb}[1]{\textcolor{black}{#1}}
\newcommand{\etal}{\emph{et al.}}

\definecolor{bgpeach}{rgb}{1.000,0.925,0.850}
\definecolor{fggray}{rgb}{0.384,0.435,0.471}

\begin{document}


\title{A theory for the stabilization of polar crystal surfaces by a
  liquid environment}

\author{Stephen J. Cox} 
\affiliation{Yusuf Hamied Department of Chemistry, University of
  Cambridge, Lensfield Road, Cambridge CB2 1EW, United Kingdom}
\email{sjc236@cam.ac.uk}

\date{\today}

\begin{abstract}
  Polar crystal surfaces play an important role in the functionality
  of many materials, and have been studied extensively over many
  decades. In this article, a theoretical framework is presented that
  extends existing theories by placing the surrounding solution
  environment on an equal footing with the crystal itself; \tcb{this
    is advantageous, e.g., when considering processes such as crystal
    growth from solution}. By considering the polar crystal as a stack
  of parallel plate capacitors \tcb{immersed in a solution
    environment}, the equilibrium adsorbed surface charge density is
  derived by minimizing the free energy of the system. In analogy to
  the well-known diverging surface energy of a polar crystal surface
  at zero temperature, for a crystal in solution it is shown that the
  ``polar catastrophe'' manifests as a diverging free energy cost to
  perturb the system from equilibrium. \tcb{Going further than
    existing theories, the present formulation predicts that
    fluctuations in the adsorbed surface charge density become
    increasingly suppressed with increasing crystal thickness.} We
  also show how, in the slab geometry often employed in both
  theoretical and computational studies of interfaces, an electric
  displacement field emerges as an electrostatic boundary condition,
  the origins of which are rooted in the slab geometry itself, rather
  than the use of periodic boundary conditions. \tcb{This aspect of
    the work provides a firmer theoretical basis for the recent
    observation that standard ``slab corrections'' fail to correctly
    describe, even qualitatively, polar crystal surfaces in solution.}
\end{abstract}

\maketitle


\section{Introduction}
\label{sec:Intro}

Crystal morphology is a determining factor in the mechanical,
rheological and catalytic properties of ionic crystals
\cite{dandekar2013engineering}. As such, there is considerable effort
to establish simple and inexpensive routes to control which facets of
a crystal are exposed at its surfaces. In this article, we will
concern ourselves exclusively with polar crystal faces, which we
describe in more detail in Sec.~\ref{subsec:Derivation}. Examples of
polar crystal surfaces include halite \hkl(111), wurtzite \hkl(0001),
and zincblende \hkl(111). Polar surfaces have been studied extensively
from a solid state physics perspective
\cite{tasker1979stability,noguera2000polar,goniakowski2008polarity},
owing to their use in microelectronic devices and the now
well-established fabrication techniques of thin metal oxide
films. Polar surfaces are also important for catalysis. For example,
the \hkl{111} facets of MgO exhibit ultrahigh catalytic activity
\cite{zhu2006efficient} for Claisen-Schmidt condensation of
benzaldehyde and acetophone, and the \hkl(0001) face of ZnO is active
in methanol \cite{cheng1983structure} decomposition.
  
What makes polar surfaces both challenging and interesting is that
basic electrostatic arguments show that their surface energy diverges
with increasing crystal thickness along a polar crystallographic
direction \cite{nosker1970polar,tasker1979stability,noguera2000polar};
this is the so-called ``polar catastrophe.'' The fact that polar
crystal surfaces are observed thus implies a ``polarity compensation
mechanism'' that overcomes the underlying instability. Known polarity
compensation mechanisms are
\cite{noguera2000polar,goniakowski2008polarity}: (i) charge transfer
via nonstoichiometric reconstruction, which amounts to an effective
transfer of ions from one side of the crystal to the other; (ii)
electronic reconstruction, where interfacial charges are modified by
partial filling of electronic interface states; and (iii) adsorption
of foreign atoms or ions that provide an appropriate amount of
compensating interfacial charge. The theoretical rationalization of
the polar catastrophe from energetic considerations of a crystal
surface in contact with a low-density vapor phase lends itself most
naturally to describing mechanisms (i) and (ii).

Polarity compensation through adsorption of ions from a surrounding
liquid environment can be considered an extreme example of mechanism
(iii), and provides a potentially simple and cost-effective route to
controlled exposure of polar crystal facets. For example, using the
molten salt synthesis route, Xu \etal \cite{xu2009syntheses} have
synthesized particles of \ce{ZnO}, \ce{MgO} and \ce{Co3O4} that
exclusively expose polar facets, and subsequent studies from Susman
\etal{} suggest that \ce{K+} and \ce{Cl-} ions potentially stabilize
the \hkl{111} facets of MgO \cite{susman2018factors}, while a complex
interplay between the molten salt and resulting crystal morphology is
evident for \ce{NiO} \cite{susman2020synthesis}. To understand the
processes that underlie polarity compensation by ion adsorption from a
surrounding liquid, it will be useful to have a theory that treats the
environment on an equal footing to the crystal. The purpose of this
article is to develop such a theoretical framework.

A second motivation is to reiterate the central message of Tasker's
seminal article \cite{tasker1979stability} on the theory of polar
surfaces: to demonstrate that the polar catastrophe has a true
physical origin and is not the result of lattice summation
techniques. Such a (re)clarification is necessary as the use of
molecular simulations---and consequently, lattice summation
techniques---has exploded since Tasker's original 1979 publication,
and the technical nature of treating long-range electrostatic
interactions under periodic boundary conditions can obfuscate physical
intuition. Similar to Tasker, then, the theory presented here will not
rely on periodic boundary conditions. Nonetheless, we will see
implications for simulations that do employ periodic boundary
conditions, and we make two observations that are useful from a
practical viewpoint when modelling, with a slab geometry, the surfaces
of polar crystals in contact with a conducting medium such as an
electrolyte:
\begin{enumerate}
\item{tinfoil Ewald approaches\tcb{, briefly described in
    Sec.~\ref{subsec:Validation},} are appropriate if we are
  interested in genuinely thin polar crystals.}
\item{An electric displacement field emerges as a boundary
  condition.}
\end{enumerate}
The second point clarifies the role of the electric displacement field
imposed in previous simulation studies
\cite{sayer2017charge,sayer2019finite,sayer2020macroscopic}. \tcb{Specifically,
  the current work emphasizes that it is conceptually incorrect to
  associate this electric displacement field with the removal of
  spurious interactions between periodic replicas of the simulation
  cell, and places the observation made in
  Ref.~\onlinecite{sayer2020macroscopic}---that standard ``slab
  corrections'' lead to qualitatively incorrect results---on a firmer
  theoretical footing.}

The rest of the article is organized as follows. In
Sec.~\ref{subsec:Derivation} we derive an expression for the
equilibrium adsorbed surface charge density at halite \hkl(111) in
contact with an electrolyte solution, as a function of crystal
thickness. We also show how the polar catastrophe manifests as a
diverging free energy cost for fluctuations away from equilibrium as
the crystal thickness increases. In Sec.~\ref{subsec:Validation}, we
show that the theoretical predictions are consistent with results from
molecular dynamics simulations that employ a tinfoil Ewald approach
to treat electrostatic interactions. In Sec.~\ref{sec:Demerge}, we
discuss the emergence of the electric displacement field in the slab
geometry, and also outline a justification for the use of tinfoil
Ewald approaches to model genuinely thin polar crystals. We summarize
our findings in Sec.~\ref{sec:Conclusions}, before providing a brief
summary of the simulation methods in Sec.~\ref{sec:Methods}. We also
provide an appendix detailing a more formal justification for the use
of tinfoil Ewald approaches.

\section{The polar catastrophe from free energy minimization}
\label{sec:Theory}

\subsection{Derivation}
\label{subsec:Derivation}

\begin{figure*}[tb]
  \includegraphics[width=12cm]{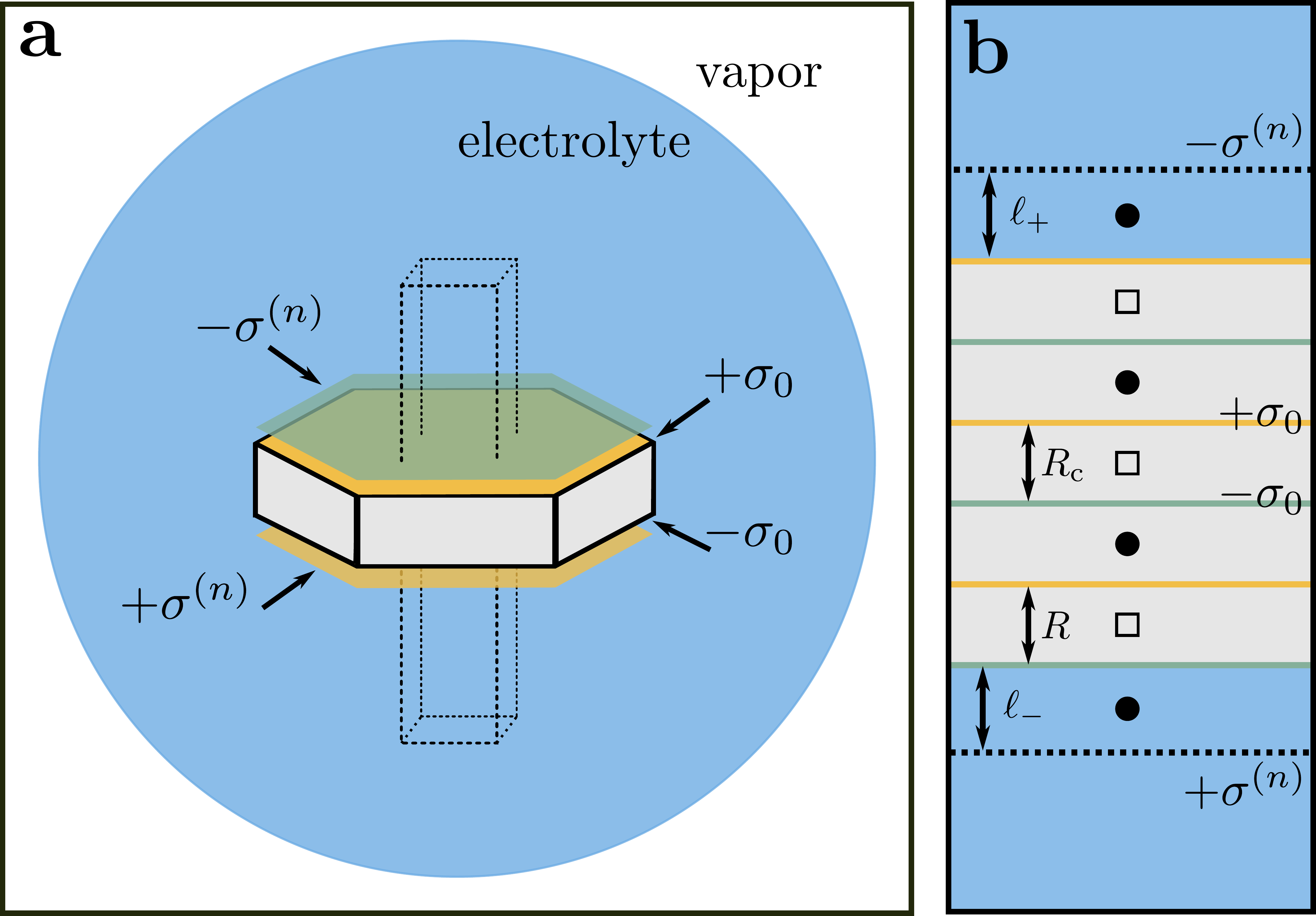}
  \caption{Schematic of a polar crystal in solution. (a) We imagine
    that a crystal exposing polar surfaces with charge density
    $\pm\sigma_0$ is immersed in an electrolyte environment. We
    suppose that anions from solution adsorb to the positive crystal
    surface, giving rise to adsorbed surface charge density
    $-\sigma^{(n)}$. Similarly, at the negative crystal surface the
    adsorbed surface charge density is $+\sigma^{(n)}$. At some
    macroscopic distance away from the crystal, there is a boundary
    between the electrolyte and its vapor phase. To investigate the
    properties of these polar surfaces, we imagine taking a cut of the
    system far from the edges of the crystal, as indicated by the
    dashed box, and ignore edge effects; this is the ``slab
    geometry.''  (b) In the slab geometry, the polar crystal is
    modeled as $n+1$ planes with alternating charge density
    $\pm\sigma_0$, separated by a distance $R$. The specific case of
    $n=5$ is shown. We will also consider cases where the central
    planes are separated by a distance $R_{\rm c}$. The planes of
    adsorbed surface charge density (dotted lines) are separated from
    the positive and negative termini by $\ell_+$ and $\ell_-$,
    respectively. In regions marked with filled circles, the electric
    field is $E = 4\pi\sigma^{(n)}$, while in those marked with empty
    squares $E = 4\pi(\sigma^{(n)}-\sigma_0)$.}
  \label{fig:schematic1}
\end{figure*}

A schematic the system we are interested is shown in
Fig.~\ref{fig:schematic1}a. Here, a crystal exposing polar facets is
immersed in an electrolyte environment. We assume that the crystal
assumes a bulk truncated structure, i.e., nonstoichiometric
reconstruction [mechanism (i)] has not taken place. Moreover, we
assume an ionic model such that electronic reconstruction [mechanism
  (ii)] is not possible. With these approximations, the exposed
crystal faces have surface charge density $\pm\sigma_0$. We now
suppose that ions from solution adsorb to the polar crystal faces with
surface charge density $\mp\sigma^{(n)}$. The reason for the ``$(n)$''
superscript will become clear shortly. To investigate these surfaces,
we imagine taking a cut of the system far from the edges of the
crystal, as indicated by the dotted box in Fig.~\ref{fig:schematic1}a,
and ignore any edge effects. This amounts to assuming that the planes
of charge span the entire plane orthogonal to the surface normal. This
``slab geometry,'' which is shown in greater detail in
Fig.~\ref{fig:schematic1}b, is similar to the textbook parallel plate
capacitor model \cite{ZangwillModernElectro}, a fact we draw upon
heavily in what follows. Despite relying on the parallel plate
capacitor model extensively, we should acknowledge that neglecting
edge effects in the slab geometry is not an innocent omission; in
Sec.~\ref{sec:Demerge} we will see that doing so ultimately gives rise
to a boundary condition that is absent in the system of interest
(Fig.~\ref{fig:schematic1}a).

Also given in Fig.~\ref{fig:schematic1}b is a more detailed
description of the structure of the polar crystal. For simplicity, we
consider halite \hkl(111)---relevant to, e.g., MgO, NiO, and NaCl---in
which $n+1$ crystal planes are separated by a distance $R$, where $n$
is an odd integer, though results can be generalized to other
geometries. Later, we will also perform ``simulation experiments'' in
which we change the separation between the central planes in the
crystal to $R_{\rm c}\neq R$. In the slab geometry, the underlying
average microscopic charge density $\rho_{\rm c}(z)$ only varies in
the direction normal to the surface, which we indicate by $z$. The
total adsorbed surface charge density is then determined by
\begin{equation}
  \label{eqn:SigmaFromMic}
  \sigma^{(n)} = \int_{{\rm int}-}\mrm{d}z\,\rho_{\rm c}(z) = -\int_{{\rm int}+}\mrm{d}z\,\rho_{\rm c}(z),
\end{equation}
where ``int$-$'' and ``int$+$'' indicate that the integrations are
performed over interfacial regions corresponding to the negatively and
positively charged crystal surfaces,
respectively.\footnote{Equation~\ref{eqn:SigmaFromMic} implicitly
  assumes that an equal number of crystal planes with $\sigma_0$ and
  $-\sigma_0$ are included in the domain of integration.} While
Eq.~\ref{eqn:SigmaFromMic} provides a direct means to obtain
$\sigma^{(n)}$ from average microscopic properties, deciding where to
locate the planes of adsorbed surface charge density is less
straightforward. Nonetheless, it seems reasonable to place the plane
with charge density $\sigma^{(n)}$ a distance $\ell_-$ from the
crystal's negatively charged surface, and that with $-\sigma^{(n)}$ a
distance $\ell_+$ from the crystal's positively charged surface, where
$\ell_-$ and $\ell_+$ both have a clear microscopic interpretation,
inasmuch as they can be related to features of $\rho_{\rm c}(z)$. In
what follows, we will in fact find an optimal \emph{apparent} length
scale entering the continuum model that has no such clear microscopic
interpretation.

Let us denote the free energy per unit area of the system as
$f(\sigma^{(n)})$. Despite discussing the system in terms of uniform
planes of charge, we have already acknowledged that the underlying
system comprises atomic and molecular entities. It is natural, then,
to partition $f$ into ``capacitive'' and ``noncapacitive''
contributions:
\begin{equation}
  \label{eqn:f}
  f(\sigma^{(n)}) = u_{\rm cap}(\sigma^{(n)}) + f_{\rm nc}(\sigma^{(n)}).
\end{equation}
The capacitive contribution, $u_{\rm cap}$, captures the energy stored
in the electric fields assuming that the system comprises uniform
planes of charge, as shown in Fig.~\ref{fig:schematic1}b. The
noncapacitive term, $f_{\rm nc}$, captures contributions from
everything else, e.g., non-electrostatic interactions, electrostatic
interactions not captured by the simple capacitor model, solvent
effects, and any entropic contributions.

To find $u_{\rm cap}$, we view the crystal in the slab geometry as
$(n+1)/2$ parallel plate capacitors, between which the electric field
is $E = 4\pi(\sigma^{(n)}-\sigma_0)$, along with $(n-1)/2$ capacitors
between which the electric field is $E = 4\pi\sigma^{(n)}$. For both
sets of capacitors, the separation between plates is $R$. (For the
moment, we consider $R_{\rm c}=R$.) In addition, the electric field
between an adsorbed plane and the surface of the crystal is also $E =
4\pi\sigma^{(n)}$. Recalling that the energy density of an electric
field is $|E|^2/8\pi$, it follows that
\begin{align}
  \label{eqn:ucap}
  u_{\rm cap} &= 2\pi(nR + 2\ell)\big(\sigma^{(n)}\big)^2 \nonumber \\
  &-2\pi(n+1)R\sigma_0\sigma^{(n)} + \pi(n+1)R\sigma_0^2,
\end{align}
where $2\ell = l_++l_-$. (Throughout this formulation, we work in a
unit system in which $4\pi\epsilon_{0} = 1$, where $\epsilon_{0}$ is
the permittivity of free space.) At equilibrium, $\sigma^{(n)}$
assumes a value that minimizes $f$:
\begin{equation}
  \label{eqn:SigmaEqGen}
  \sigma^{(n,\rm eq)} = \frac{(n+1)R\sigma_0 - f^\prime_{\rm nc}/2\pi}{2nR + 4\ell},
\end{equation}
where the prime indicates a partial derivative with respect to
$\sigma^{(n)}$. Equation~\ref{eqn:SigmaEqGen} is a central result of
this article. While noncapacitive contributions are irrelevant as
$n\to\infty$, their effects become increasingly important as the
thickness of the crystal decreases.

Although formally exact, analysis of Eq.~\ref{eqn:SigmaEqGen} is
complicated by the presence of the noncapacitive contributions. To
make exploratory progress, we simply postulate that
\begin{equation}
  \label{eqn:fnc-phenom}
  f_{\rm nc}(\sigma^{(n)}) = 4\pi a_{\rm nc}\big(\sigma^{(n)}\big)^2.
\end{equation}
The equilibrium adsorbed surface charge density then reads
\begin{equation}
  \label{eqn:SigmaEqa}
  \sigma^{(n,\rm eq)} = \frac{(n+1)R\sigma_0}{2nR + 4(\ell+a_{\rm nc})}.
\end{equation}
Within the phenomenological model specified by
Eq.~\ref{eqn:fnc-phenom}, one can view the effects of noncapacitive
contributions as modifying the apparent length scale associated with
ion adsorption, from $\ell$ to $\ell+a_{\rm nc}$, in the continuum
representation of the system. A similar expression for $\sigma^{(n,\rm
  eq)}$ was derived previously by Hu \cite{hu2021comment} for the slab
geometry under periodic boundary conditions, within the framework of
symmetry-preserving mean field theory
\cite{pan2019analytic,hu2014symmetry}. Within that framework, an
effective length scale is rationalized by considering the slowly
varying components of $\rho_{\rm c}$, whose structural features need
not coincide with those of the full average microscopic charge
density. We will postpone further discussion of the effective length
scale $\ell+a_{\rm nc}$ to Sec.~\ref{subsec:Validation} and
Appendix~\ref{Appendix}. Importantly, the physical interpretation of
Eq.~\ref{eqn:SigmaEqa} is somewhat different from that of
Ref.~\onlinecite{hu2021comment}, where $\sigma^{(n)} = \sigma^{(n,\rm
  eq)}$ was attributed to the presence of a spurious electric field
arising from the use of tinfoil Ewald sums. In contrast, in this
study Eq.~\ref{eqn:SigmaEqa} has been derived for a nonperiodic slab
geometry; the driving force for ion adsorption is a significant
reduction in the capacitive energy stored in the system upon ion
adsorption.

To understand the origin of the polar catastrophe from this free
energy perspective, we consider the reversible work required to change
the adsorbed surface charge density by an amount $\delta\sigma$ from
its equilibrium value,
\begin{equation}
  \nonumber
  f(\sigma^{(n,\rm eq)}+\delta\sigma) = 
  f(\sigma^{(n,\rm eq)}) + \frac{1}{2}f^{\prime\prime}(\sigma^{n,\rm eq})(\delta\sigma)^2
  + \mathcal{O}\big((\delta\sigma)^3\big).
\end{equation}
Neglecting terms higher than second order in $\delta\sigma$, and
combining with Eqs.~\ref{eqn:f}, \ref{eqn:ucap}
and~\ref{eqn:fnc-phenom} gives
\begin{equation}
  \label{eqn:fluct}
  f(\sigma^{(n,\rm eq)}+\delta\sigma) - f(\sigma^{(n,\rm eq)}) = 2\pi\big(nR + 2(\ell+a_{\rm nc})\big)(\delta\sigma)^2.
\end{equation}
Equation~\ref{eqn:fluct} is another key result of this study; it is
analogous to the familiar diverging surface energy of an
unreconstructed polar crystal in contact with vacuum
\cite{tasker1979stability,noguera2000polar}. We see that the
reversible work required to change the adsorbed surface charge density
from $\sigma^{(n,\rm eq)}$ by any finite amount diverges linearly with
the crystal thickness $nR$.

\tcb{Despite the clear analogy to the well-known diverging surface
  energy of polar crystal surfaces in vacuum, Eq.~\ref{eqn:fluct}
  contains greater information. This can be seen more clearly by
  considering the} behavior of the fluctuations of the total charge
$Q$ within a probe area $A^{(n)}$ at the surface of the crystal. At
equilibrium, the total adsorbed charge within such an area is
$Q^{(n,\rm eq)} = A^{(n)}\sigma^{(n,\rm eq)}$. Eq.~\ref{eqn:fluct}
states that the area required to observe $\delta Q = Q - Q^{(n,\rm
  eq)} \approx e$ with appreciable probability, i.e., $\beta
A^{(n)}[f(\sigma^{(n,\rm eq)}+\delta\sigma) - f(\sigma^{(n,\rm
    eq)})]\approx 1$, is,
\begin{equation}
  A^{(n)} \approx 2\pi\beta e^2\big(nR + 2(\ell+a_{\rm nc})\big),
\end{equation}
where $e$ is the unit of elementary charge, and $\beta = 1/k_{\rm B}T$
($k_{\rm B}$ is Boltzmann's constant, and $T$ is the temperature). As
the thickness $n$ of the crystal increases, ever increasing probe
areas are required to observe appreciable fluctuations in the adsorbed
surface charge. \tcb{The theory presented here not only reveals the
  strong constraint placed on the average adsorbed surface charge
  density at polar crystal surfaces, but also that fluctuations at the
  interface are strongly suppressed.}

\subsection{Validation of the theory with molecular simulations}
\label{subsec:Validation}

Equation~\ref{eqn:SigmaEqa} provides an expression for the equilibrium
adsorbed surface charge density at polar halite \hkl(111) surfaces in
contact with an electrolyte solution, assuming an ionic model for the
crystal. As such, Eq.~\ref{eqn:SigmaEqa} lends itself naturally to
validation with molecular simulations. But two comments are in
order. The first concerns the treatment of long-range electrostatic
interactions
under periodic boundary conditions\tcb{, for which lattice summation
  techniques are often employed. For Ewald summation, which underlies
  many of the most popular techniques used in molecular simulations,
  the expression for the Coulomb energy for a set of charges $\{q_i\}$
  with positions $\{\mbf{r}_i\}$ reads:}
\begin{align}
  U_{\rm C} &= \frac{1}{2}\sum_{\rm\mbf{b}}\sideset{}{^\prime}\sum_{i,j} q_iq_j\phi_{\rm SR}(\mbf{r}_i-\mbf{r}_j-\mbf{b}) \nonumber \\
  &+ \frac{1}{2\Omega}\sum_{\mbf{k}\neq\mbf{0}}|\tilde{\rho}(\mbf{k})|^2\tilde{\phi}_{\rm LR}(\mbf{k}) - \frac{1}{2}\sum_{i}q_i^2\phi_{\rm LR}(\mbf{0}) \nonumber \\
  &+ \frac{1}{2\Omega}\mbf{M}\cdot\mbf{J}\cdot\mbf{M}.\label{eqn:Ucoul}
\end{align}
\tcb{In Eq.~\ref{eqn:Ucoul}, $\mbf{b}$ is a lattice vector, and the
  prime indicates that the term $i=j$ is omitted when
  $\mbf{b}=\mbf{0}$. The volume of the simulation cell is $\Omega$,
  and $\tilde{\rho}(\mbf{k})$ is the Fourier transform of the charge
  density $\rho(\mbf{r}) = \sum_iq_i\delta(\mbf{r}-\mbf{r}_i)$. The
  potentials $\phi_{\rm SR}(\mbf{r})$ and $\phi_{\rm LR}(\mbf{r})$ are
  defined by a splitting of the Coulomb potential into short- and
  long-ranged contributions, $1/|\mbf{r}| = \phi_{\rm SR}(\mbf{r}) +
  \phi_{\rm LR}(\mbf{r})$, and $\tilde{\phi}_{\rm LR}$ is the Fourier
  transform of the long-ranged part. Most important for the current
  discussion is the final ``surface term,'' which depends upon the
  total dipole moment of the simulation cell, $\mbf{M}$, and the
  depolarization tensor, $\mbf{J}$. This surface term is determined by
  the specified summation order of the lattice sum, and for a
  spherical summation order it is equal to
  $2\pi|\mbf{M}|^2/(2\epsilon^\prime+1)$, where $\epsilon^\prime$ is
  the dielectric constant of a surrounding medium ``at infinity.''
  Tinfoil Ewald approaches take this surrounding medium to be a
  perfect conductor, $\epsilon^\prime = \infty$, and amounts to
  ignoring the surface term. The above is not intended as a detailed
  discussion of Ewald summation, for which there is an extensive
  literature
  \cite{LPS1,LPS2,smith1981electrostatic,de1986computer,neumann1983dipole,neumann1983calculation1,neumann1983calculation2,neumann1984computer,redlack1975coulombic,kantorovich1999coulomb,smith2008electrostatic}.
  The reader is referred to
  Ref.~\onlinecite{ballenegger2014communication} for a particularly
  clear discussion on the topic.}

\tcb{For systems in the slab geometry under periodic boundary
  conditions employing tinfoil Ewald, where the slab is surrounded on
  either side by vacuum, it has long been recognized that if a net
  dipole $M_z$ exists along $z$, then there is a finite electric field
  in the vacuum region. It is therefore common to employ ``slab
  corrections'' to mitigate possible spurious interactions between
  periodic images. In classical simulations of liquid-solid
  interfaces, the standard choice is the method of Yeh and Berkowitz
  \cite{YehBerkowitz1999sjc}, which includes a surface term $2\pi
  M_z^2/\Omega$. A similar expression exists in the surface science
  community (which typically uses periodic density functional theory
  calculations)\cite{neugebauer1992adsorbate,PhysRevB.59.12301}, where
  it is known as the ``dipole correction.''} In
Ref.~\onlinecite{sayer2020macroscopic}, however, it was demonstrated
that typical slab correction schemes impose vanishing adsorbed surface
charge density; more generally, it was found that an electric
displacement field directly determines the adsorbed surface charge,
irrespective of the slab's thickness. Such a scenario is clearly
incompatible with $\sigma^{(n,\rm eq)}$ predicted by
Eq.~\ref{eqn:SigmaEqa}. In contrast, simulations performed with
tinfoil Ewald approaches were found to give adsorbed surface charge
densities that increase with crystal thickness. We will therefore
compare the predictions of the theory presented above to simulations
that use a tinfoil Ewald approach; a theoretical justification for
doing so will be given in Sec.~\ref{sec:Demerge}\tcb{, where it will
  also become clear why standard slab corrections fail}.

\begin{figure}[tb]
  \includegraphics[width=8cm]{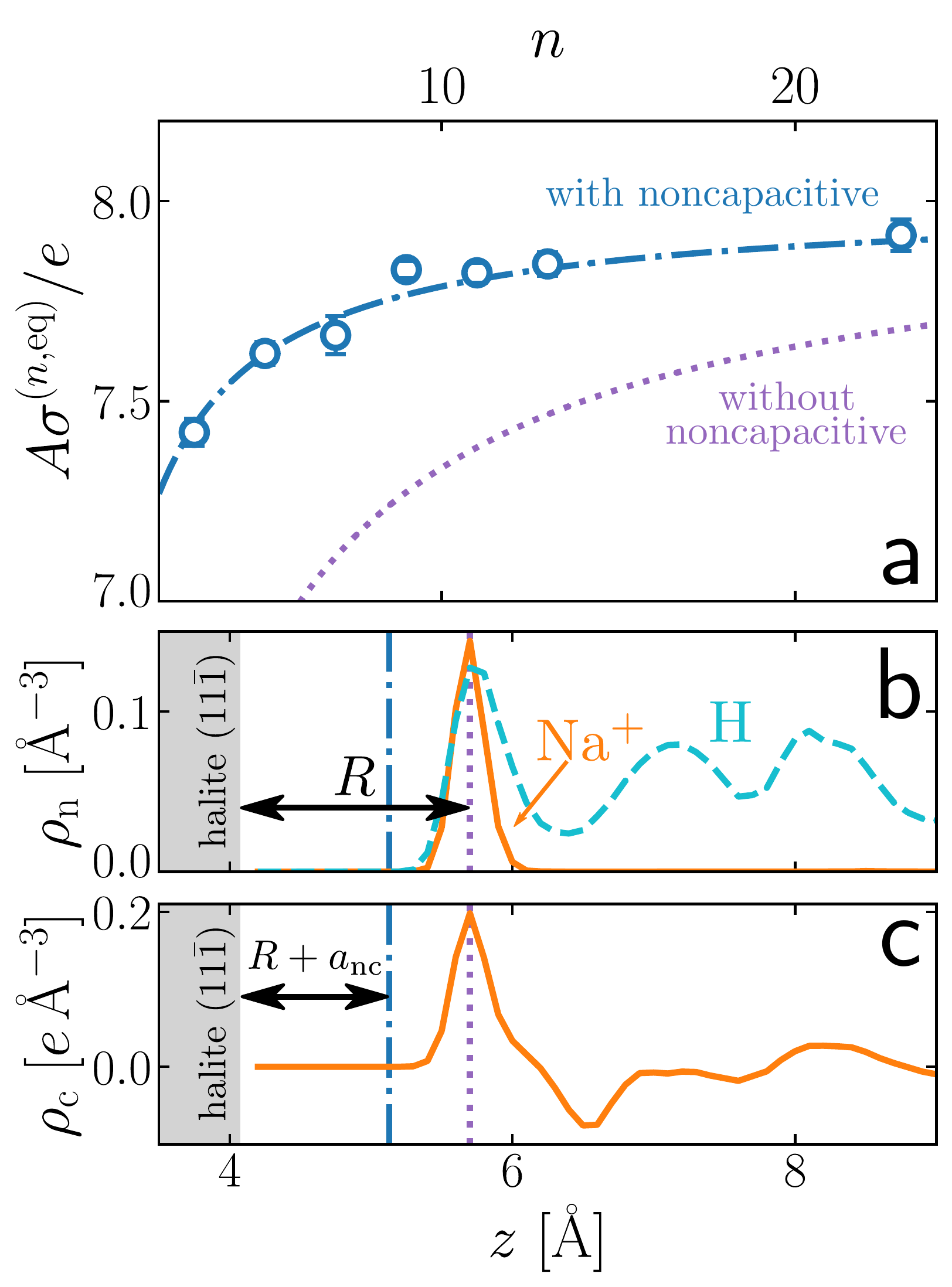}
  \caption{Theoretical predictions and molecular simulations paint a
    consistent picture for ion adsorption at polar surfaces. (a)
    Symbols show $\sigma^{(n,\rm eq)}$ obtained from 10\,ns
    simulations of a halite slab exposing its \hkl(111) and \hkl(11-1)
    surfaces to aqueous electrolyte solution. The dot-dashed line
    indicates a best fit of the theoretical prediction
    (Eq.~\ref{eqn:SigmaEqa}) to the simulation data, with $\ell = R$
    and $a_{\rm nc}=-0.568$\,\AA, where $a_{\rm nc}$ is the only
    fitting parameter. A model that ignores noncapacitive
    contributions ($a_{\rm nc}=0$) gives a poor description of the
    simulation data. $A\approx 15.95\times 13.82$\,\AA$^2$ is the
    cross-sectional area of the simulation cell. Error bars indicate
    95\,\% confidence intervals, estimated by splitting trajectories
    into five samples. (b) Average number density profiles $\rho_{\rm
      n}(z)$ for \ce{Na+} ions (solid orange line) and water hydrogen
    atoms (dashed blue line) above halite \hkl(11-1). (c) Average
    charge density $\rho_{\rm c}(z)$ above halite \hkl(11-1). In both
    (c) and (d), the region occupied by the crystal is indicated by
    the shaded gray region. The vertical purple dotted line indicates
    the separation $R$ from the crystal's surface, which aligns well
    with the first peaks in $\rho_{\rm n}(z)$ and $\rho_{\rm
      c}(z)$. The vertical blue dot-dashed line indicates the
    separation $R+a_{\rm nc}$, which lacks a clear microscopic
    interpretation.}
  \label{fig:SigmasNaClOnly}
\end{figure}

The second comment concerns the parameters that determine
$\sigma^{(n,\rm eq)}$. Here, we will consider crystals where the
positions of the ions have been clamped, and whose charges do not
fluctuate; $\sigma_0$ and $R$ are then straightforward input
parameters. The extent to which agreement between theory and
simulation is observed is an indicator of how faithfully the theory
presented in this article describes the ionic model of polar
surfaces. Moreover, we will consider the specific example of NaCl
\hkl(111) in contact with an aqueous NaCl solution. We then anticipate
that ions in direct contact with the crystal will form a plane
separated from the crystal by a distance $\ell\approx R$. The only
remaining unknown parameter is $a_{\rm nc}$, the length scale
associated with noncapacitive contributions.

In a first validation step, we compare $\sigma^{(n,\rm eq)}$ obtained
from molecular simulations of a concentrated aqueous NaCl solution in
contact with halite \hkl(111), where $a_{\rm nc}$ is simply treated as
a free fitting parameter. As seen in Fig.~\ref{fig:SigmasNaClOnly}a,
Eq.~\ref{eqn:SigmaEqa} with $\ell = R = 1.628$\,\AA{} and $a_{\rm
  nc}=-0.568$\,\AA{} captures the trend seen in the molecular
simulations very well. To determine the significance of noncapacitive
contributions, it is instructive to compare the effective length scale
$\ell+a_{\rm nc} = 1.06$\,\AA{} in the continuum model to real length
scales in the system. To that end, Fig.~\ref{fig:SigmasNaClOnly}b
shows the number density profile of \ce{Na+} ions at halite \hkl(11-1)
with $n=5$. As expected, the plane separated from halite \hkl(11-1) by
$R$ coincides with the plane of \ce{Na+} directly adsorbed to the
surface. In contrast, the plane at $\ell+a_{\rm nc}$ from \hkl(11-1)
is located in a region that cannot be readily associated with ion
adsorption; it does not appear that $a_{\rm nc}$ simply captures
effects due to thermal fluctuations.

The catch-all nature of $f_{\rm nc}$, and the \emph{ad hoc} assertion
of its quadratic form (Eq.~\ref{eqn:fnc-phenom}), makes direct
physical interpretation of the value of $a_{\rm nc}$
challenging. Recent work suggests that dielectric boundaries for
systems with water as solvent are closely associated with water's
hydrogen density \cite{cox2021quadrupole,cox2022dielectric}. Also
plotted in Fig.~\ref{fig:SigmasNaClOnly}b, therefore, is the number
density profile of water's hydrogen atoms. While some hydrogen density
is found marginally closer to halite \hkl(11-1) than that of the
\ce{Na+} ions, it is difficult to draw any firm conclusions on a
possible relationship between $a_{\rm nc}$ and the solvent. More
likely is that noncapacitive contributions account for the rather
drastic approximation of collapsing the entire adsorbed surface charge
density into a single plane. As can be seen in
Fig.~\ref{fig:SigmasNaClOnly}c, $\rho_{\rm c}(z)$ exhibits pronounced
structure beyond the first peak at $\ell = R$. In
Sec.~\ref{sec:Demerge}, we will see that the effective length scale
$\ell+a_{\rm nc}$ in the continuum model is that which equalizes the
electrostatic potential in the solution on either side of the crystal,
a condition that is automatically satisfied by molecular simulations
of a crystal slab immersed in an electrolyte solution, in which
tinfoil Ewald approaches are used (see Appendix~\ref{Appendix}).

Despite the lack of a simple physical interpretation, the importance
of $f_{\rm nc}$ can be readily seen by plotting Eq.~\ref{eqn:SigmaEqa}
with $a_{\rm nc}=0$, as shown in
Fig.~\ref{fig:SigmasNaClOnly}a. Clearly, a model that only considers
capacitive contributions in which the system is approximated as planes
of uniform charge density poorly describes the simulation data, with
predicted adsorbed surface charge densities much lower than that
observed.

A possible cause for concern is that simply treating $a_{\rm nc}$ as
free fitting parameter amounts to nothing more than a convenient
fix. Challenges in interpreting the value of $a_{\rm nc}$
notwithstanding, it is reasonable to assume that noncapacitive
contributions largely arise from effects that are localized to the
solution and interface regions. While changes to the structure of the
crystal in regions far from the interface will affect $u_{\rm cap}$,
we do not anticipate significant impact on the coefficient $a_{\rm
  nc}$. In a second validation step, we therefore perform a
``simulation experiment'' in which the separation between the
crystal's central planes is varied (i.e., $R_{\rm c}\neq R$). The
manner in which $\sigma^{(n,\rm eq)}$ is affected by $R_{\rm c}\neq R$
depends on whether the central planes bound a region with $E =
4\pi\sigma^{(n)}$ [$(n+1)/2$ is even] or $E =
4\pi(\sigma^{(n)}-\sigma_0)$ [$(n+1)/2$ is odd]. Following the same
approach leading to Eq.~\ref{eqn:SigmaEqa} gives
\begin{widetext}
  \begin{subnumcases}{\sigma^{(n,\rm eq)} =}
    \frac{\left[(n-1)R + 2R_{\rm c}\right]\sigma_0}{\left[(2n-2)R + 2R_{\rm c}+4(\ell+a_{\rm nc})\right]} & [\text{$(n+1)/2$ is odd}], \label{eqn:Rc-odd} \\[7pt]
    \frac{(n+1)R\sigma_0}{\left[(2n-2)R + 2R_{\rm c}+4(\ell+a_{\rm nc})\right]} & [\text{$(n+1)/2$ is even}]. \label{eqn:Rc-even}
  \end{subnumcases}
\end{widetext}

\begin{figure}[tb]
  \includegraphics[width=8cm]{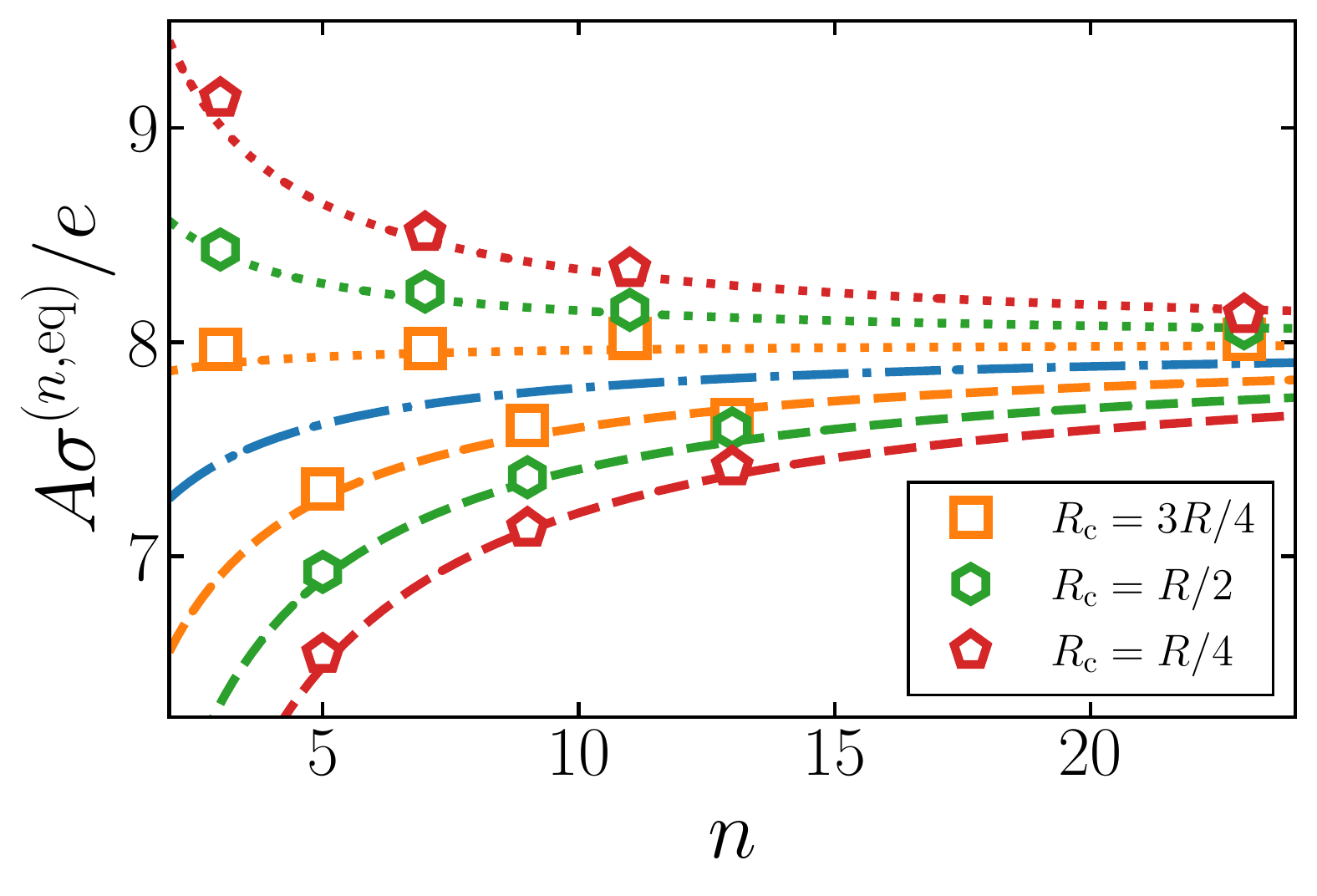}
  \caption{``Simulation experiments'' further validate the
    theory. Symbols show $\sigma^{(n,\rm eq)}$ obtained from
    simulations with $R_{\rm c}\neq R$, as indicated in the
    legend. Dashed and dotted lines indicate theoretical predictions
    described by Eqs.~\ref{eqn:Rc-odd} and~\ref{eqn:Rc-even},
    respectively, with $a_{\rm nc}=-0.568$\,\AA{} (no further
    fitting). The good agreement between simulation and theory
    suggests that $a_{\rm nc}$ is insensitive to structural changes
    away from the interface. The blue dot-dashed line is the same as
    Fig.~\ref{fig:SigmasNaClOnly}a, i.e., $R_{\rm c}=R$.}
  \label{fig:Sigmas}
\end{figure}

Results obtained from simulations with $R_{\rm c} = 3R/4$, $R/2$ and
$R/4$ are presented in Fig.~\ref{fig:Sigmas}, along with the
theoretical predictions described by Eqs.~\ref{eqn:Rc-odd}
and~\ref{eqn:Rc-even}, in which no further fitting has been
performed. That is, the same value $a_{\rm nc}=-0.568$\,\AA{} has been
used. The good agreement between the theoretical predictions and the
simulation data lends support to the notion that fitting $a_{\rm nc}$
captures genuine noncapacitive effects that are localized to the
solution and interface regions.

Results presented so far demonstrate consistency between the theory
presented in Sec.~\ref{subsec:Derivation} for the equilibrium adsorbed
surface charge density at halite \hkl(111), and molecular simulations
that use a tinfoil Ewald approach. In Sec.~\ref{sec:Demerge}, we will
discuss why it is appropriate to compare $\sigma^{(n,\rm eq)}$ for the
system of interest (Fig.~\ref{fig:schematic1}a) to simulations that
employ the slab geometry under periodic boundary conditions. In doing
so, we will also shed light on the role of the electric displacement
field in the slab geometry
\cite{zhang2016finite,zhang2018communication,sayer2019finite,sayer2019stabilization,zhang2020modelling,sayer2020macroscopic}.

\section{The emergence of the electric displacement field in a slab geometry}
\label{sec:Demerge}

\begin{figure*}[tb]
  \includegraphics[width=12cm]{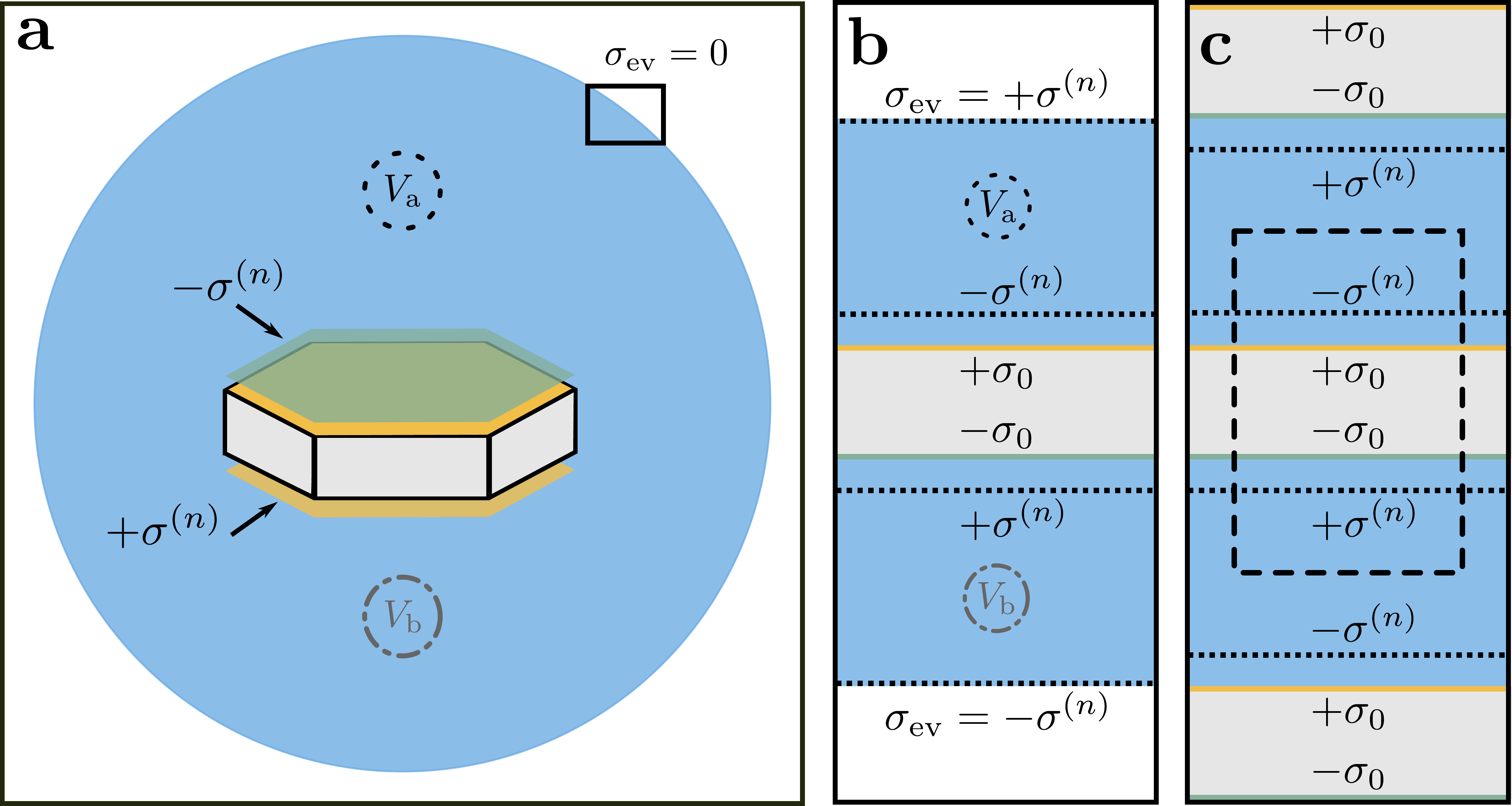}
  \caption{Emergence of the electric displacement field in a slab
    geometry. (a) The adsorbed surface charge densities $\sigma^{(n)}$
    and $-\sigma^{(n)}$ comprise ions originating from the same pool
    of electrolyte, which remains electroneutral. Both the surface
    charge density at the electrolyte-vapor interface, $\sigma_{\rm
      ev}$, and the electric field in vacuum, $\mbf{E}_{\rm v}$, are
    zero. As the electrolyte is conducting, its interior is
    equipotential, $V_{\rm a}=V_{\rm b}$. In contrast, the slab
    geometry creates two distinct regions of electrolyte either side
    of the crystal slab, as shown in (b). Unlike
    Fig.~\ref{fig:schematic1}b, we have now included the vapor phase,
    and only the outermost crystal planes are shown explicitly. As
    electroneutrality must be preserved in each region separately,
    $\sigma^{(n)} = \sigma^{(n,\rm eq)}$ implies a polarized
    electrolyte with a finite displacement field,
    $\hat{\mbf{z}}\cdot\mbf{D}_{\rm e} = 4\pi\sigma^{(n,\rm
      eq)}$. While this satisfies $V_{\rm a}=V_{\rm b}$, the electric
    field in vapor is now finite, $\hat{\mbf{z}}\cdot\mbf{E}_{\rm
      v}=4\pi\sigma^{(n,\rm eq)}$. If $\sigma^{(n)} =
    \hat{\mbf{z}}\cdot\mbf{E}_{\rm v} = 0$, then $V_{\rm a}-V_{\rm
      b}=2\pi(n+1)R\sigma_0$. Panel (c) depicts the system under a
    \emph{periodic} slab geometry, without the vapor region. While for
    $\sigma^{(n)} = \sigma^{(n,\rm eq)}$ it is still the case that
    $\hat{\mbf{z}}\cdot\mbf{D}_{\rm e} = 4\pi\sigma^{(n,\rm eq)}$, the
    electric field in the electrolyte \tcb{vanishes,} $\mbf{E}_{\rm e}
    = \mbf{0}$, and $\mbf{D}_{\rm e}\cdot\mbf{E}_{\rm e} =
    0$. Moreover, $V_{\rm a}=V_{\rm b}$.}
  \label{fig:schematic2}
\end{figure*}

Recall the procedure employed in Sec.~\ref{subsec:Derivation} to
investigate the polar surfaces of a crystal in solution. First, a
particle of finite size, with polar surfaces exposed, was immersed in
an electrolyte, and ions from solution were imagined to adsorb to its
surfaces. Then, we constructed the slab geometry by taking a cut of
the system far from the edges of the crystal, and considered the free
energy per unit area. Why did we adopt this procedure, rather than
starting immediately with the slab geometry? Consider again the system
of interest, which, to avoid notational clutter, is shown again in
Fig.~\ref{fig:schematic2}a. As the particle has finite size, the ions
adsorbed to the positive and negative crystal faces originate from the
same pool of ions in the surrounding electrolyte solution. The
remaining electrolyte solution is electroneutral. As the electrolyte
solution is conducting, the total electric field in its interior
vanishes, $\mbf{E}_{\rm e} = \mbf{0}$. Now consider the vapor
region. We will assume that the vapor pressure is sufficiently low
that this region can be approximated as vacuum. In the absence of an
external field, the total electric field there too vanishes,
$\mbf{E}_{\rm v} = \mbf{0}$. It then follows that the surface charge
density at the electrolyte-vapor interface is zero, i.e., $\sigma_{\rm
  ev} = \hat{\mbf{r}}\cdot{(\mbf{E}_{\rm v}-\mbf{E}_{\rm e})}=0$,
where $\hat{\mbf{r}}$ is the outward normal to the electrolyte
surface. This scenario is possible with the procedure employed in
Sec.~\ref{subsec:Derivation}, as the electrolyte solution remains
uncharged. Moreover, there is a clear physical pathway by which the
ions have adsorbed from solution to the surface.

Imagine that we instead start directly with the slab geometry. As the
crystal spans the entire plane orthogonal to the surface normal, there
are now two distinct regions of electrolyte. Upon ion adsorption,
electroneutrality must be preserved in each region separately (the
ions cannot pass through the crystal). As the electrolyte is a
conductor, the remaining charge in solution must be located at the
electrolyte-vapor boundary, $|\sigma_{\rm ev}|=\sigma^{(n)}$, as
indicated in Fig.~\ref{fig:schematic2}b. As before, $\mbf{E}_{\rm e} =
\mbf{0}$, but the electrolyte is now polarized, with uniform
polarization $\hat{\mbf{z}}\cdot\mbf{P}_{\rm e} = \sigma^{(n)}$. The
electric displacement field in the electrolyte then satisfies
\begin{equation}
  \label{eqn:De-slab}
  \hat{\mbf{z}}\cdot\mbf{D}_{\rm e}=\hat{\mbf{z}}\cdot(\mbf{E}_{\rm
    e}+4\pi\mbf{P}_{\rm e}) = 4\pi\sigma^{(n)}.
\end{equation}
In the vapor, we assume that atomic density is sufficiently low that
$\mbf{P}_{\rm v}=\mbf{0}$. Matching the electric displacement field
across the boundaries leads us to conclude that
\begin{equation}
  \label{eqn:Ev-slab}
  \hat{\mbf{z}}\cdot\mbf{E}_{\rm v}=\hat{\mbf{z}}\cdot(\mbf{D}_{\rm
    v}-4\pi\mbf{P}_{\rm v}) = 4\pi\sigma^{(n)}.
\end{equation}
This result for $\mbf{E}_{\rm v}$ in the slab geometry is most severe.
Integrating the electrostatic energy density $|\mbf{E}_{\rm
  v}|^2/8\pi$ over the macroscopic volume occupied by the vapor phase
indicates a prohibitively expensive energy cost for $\sigma^{(n)}\neq
0$. Another route to arrive at the conclusion $\sigma^{(n)}=0$
\emph{when starting in the slab geometry} is that, as the electric
field external to the parallel plate capacitors vanishes, there is no
driving force for ion adsorption.

The slab geometry, despite widespread use in both theoretical and
computer simulation studies, does not correspond to a physical
reality. Consider the electrostatic potential in the regions of the
electrolyte indicated by the dotted and dot-dashed circles in
Fig.~\ref{fig:schematic2}a, which we denote $V_{\rm a}$ and $V_{\rm
  b}$, respectively. As the electrolyte is conducting, it follows
immediately that $V_{\rm a}=V_{\rm b}$. In the slab geometry, the
crystal provides an insulating layer between the two distinct regions
of electrolyte. For $\sigma^{(n)} = 0$, the electrostatic potential
difference across the crystal is,
\begin{equation}
  V_{\rm a}-V_{\rm b} = 2\pi(n+1)R\sigma_0.
\end{equation}
Even for $n=1$, this electrostatic potential difference is
substantial: $V_{\rm a}-V_{\rm b} \approx 20$\,V.
In contrast, for $\sigma^{(n)}=\sigma^{(n,\rm eq)}$
(Eq.~\ref{eqn:SigmaEqa}),
\begin{equation}
  V_{\rm a}-V_{\rm b} = 0,
\end{equation}
where, as discussed in Appendix~\ref{Appendix}, we have treated the
adsorbed planes as if they are separated from the crystal's surfaces
by the effective distance $\ell+a_{\rm nc}$. The slab geometry
fundamentally breaks the system: for $\sigma^{(n)}=0$, $\mbf{E}_{\rm
  v}=\mbf{0}$ is accurately captured, but there exists an erroneous
potential difference across the crystal; for
$\sigma^{(n)}=\sigma^{(n,\rm eq)}$, the potential difference across
the crystal is correct, but at the cost of introducing a finite field
in the vapor phase, $\hat{\mbf{z}}\cdot\mbf{E}_{\rm
  v}=4\pi\sigma^{(n,\rm eq)}$. This is a consequence of the finite
electric displacement field in the electrolyte,
$\hat{\mbf{z}}\cdot\mbf{D}_{\rm e} = 4\pi\sigma^{(n,\rm eq)}$
(Eq.~\ref{eqn:De-slab}), that emerges in the slab geometry.

What are the implications for molecular simulations? Let us consider
cases where periodic boundary conditions are used in all three
Cartesian directions, restricting ourselves to cases where periodic
replicas of the crystal slab sandwich the electrolyte solution such
that there is no vapor region, as shown in
Fig.~\ref{fig:schematic2}c. Similar to the nonperiodic slab geometry,
Fig.~\ref{fig:schematic2}b, the electrolyte is polarized,
$\hat{\mbf{z}}\cdot\mbf{P}_{\rm e}=\sigma^{(n)}$, and there is a
finite electric displacement field in the electrolyte,
$\hat{\mbf{z}}\cdot\mbf{D}_{\rm e}=4\pi\sigma^{(n)}$. For simulations
that employ tinfoil Ewald approaches to compute electrostatic
interactions, there is no electrostatic potential difference across
the simulation cell. As $\mbf{E}_{\rm e}=\mbf{0}$, it then follows
that $V_{\rm a}-V_{\rm b} = 0$. (In Appendix~\ref{Appendix}, it is
shown explicitly that $\sigma^{(n,\rm eq)}$ is the adsorbed surface
charge density that enforces $\mbf{E}_{\rm e}=\mbf{0}$ when using
tinfoil Ewald approaches.) By not including a vapor region between
periodic images, the electrostatic energy is well behaved; as
$\mbf{D}_{\rm e}\cdot\mbf{E}_{\rm e}= 0$, we can extend the
electrolyte region between periodic replicas without incurring any
cost in electrostatic energy. Simulations using tinfoil Ewald sums
thus appear to give a faithful representation of polar surfaces in
solution---in the sense of taking a cut of the system of interest as
shown in Fig.~\ref{fig:schematic1}a, albeit with a finite polarization
and electric displacement field in the electrolyte, the effects of
which, however, seem largely benign.

Implicit in the preceding discussion is that the polarization is
treated in an itinerant fashion---the ions are included in its
definition---which has been shown
\cite{caillol1994comments,cox2019finite} to satisfy key statistical
mechanical properties for electrolyte solutions, such as the
Stillinger-Lovett sum rules
\cite{stillinger1968ion,stillinger1968general}. Any discussion on the
multivaluedness of polarization under periodic boundary conditions has
also been neglected, along with implications of whether the crystal or
the electrolyte straddles the boundaries of the simulation cell. The
reader is referred to
Refs.~\onlinecite{zhang2016finite,sprik2018finite,sayer2019finite} for
a discussion of these issues in the context of molecular dynamics
simulations.

As already mentioned, Ref.~\onlinecite{sayer2020macroscopic}
demonstrated that imposing an electric displacement field directly
determines the adsorbed surface charge density. Theoretical discussion
in that work, however, was brief, and what little there was relied
heavily upon the framework provided by the ``finite field approach''
of Zhang and Sprik
\cite{zhang2016computing1,zhang2016finite,zhang2020modelling}. The
picture to emerge from the finite field approach is that of a set of
``virtual electrodes'' that either controls the electrostatic
potential difference across the simulation cell (``constant $E$'') or
controls the electric displacement field at the cell boundaries
(``constant $D$''). As the finite field approach was formulated in the
context of periodic boundary conditions, it can be challenging to
disentangle its deeper significance from issues concerning conditional
convergence of lattice sums and unwanted interactions between periodic
images. Moreover, the picture of virtual electrodes can give the
impression that one is forcing the system adopt a particular adsorbed
surface charge density with an arbitrary field. In this article, the
status of the electric displacement field as a control variable has
been clarified: it should be viewed as matching an electrostatic
boundary condition that emerges in the slab geometry---even in the
absence of periodic boundary conditions---rather than an \emph{ad hoc}
external electric field that is applied to the system. In
Ref.~\onlinecite{sayer2020macroscopic}, the surfaces of
macroscopically thick crystals in contact with an electrolyte solution
were effectively simulated by imposing $\hat{\mbf{z}}\cdot\mbf{D}_{\rm
  e} = 4\pi\sigma^{(\infty,\rm eq)}$, using the finite field
approach. Similar control of the adsorbed surface charge with an
electric displacement field has also been observed to work away from
equilibrium \cite{dufils2021computational}.

\tcb{To end this discussion, it is worth considering standard slab
  correction schemes again
  \cite{YehBerkowitz1999sjc,neugebauer1992adsorbate,PhysRevB.59.12301},
  which typically introduce a vacuum region between periodic replicas
  along $z$, and remove the electric field in the vacuum region by
  enforcing $\hat{\mbf{z}}\cdot\mbf{D}_{\rm v}=0$. While such a line
  of attack seems reasonable from the perspective of removing
  interactions between periodic images, it presupposes that the true
  physical scenario corresponds to an isolated slab in contact with
  the liquid (e.g., Fig.~\ref{fig:schematic2}b). Slabs that completely
  span the plane orthogonal to the surface normal, however, do not
  exist; this fact cannot be ignored when modeling the surfaces of
  polar crystals.}

\section{Conclusions}
\label{sec:Conclusions}

A theoretical framework to describe polarity compensation arising from
the solution environment has been presented in which the free energy
of the system has been separated into ``capacitive'' and
``noncapacitive'' contributions. For crystals that are thin along a
polar crystallographic direction, noncapacitive contributions were
found to play a significant role. As the thickness of the crystal
increases, the capacitive contribution dominates, and the adsorbed
surface charge density at equilibrium agrees with known expressions
for the required charge imbalance at polar surfaces
\cite{tasker1979stability,noguera2000polar}.

The theory presented here only attempts to capture the net adsorbed
charge at each polar crystal surface, and says nothing about the
complex structural details of the solution near the interface. A major
source of the noncapactive contributions likely arises from the
drastic approximation of assuming that all adsorbed charge lies in a
single plane. In addition, the theory assumes an ionic model for the
crystal. While the severity of such an approximation will depend on
the system under investigation \cite{noguera2000polar}, it may
nonetheless provide a useful starting point for more sophisticated
theoretical approaches.

Like previous theoretical treatments for polar crystal surfaces in
contact with vacuum \cite{tasker1979stability,noguera2000polar}, the
presented theory relies heavily on modeling the crystal as a stack of
parallel plate capacitors. It was shown that in such a slab geometry,
an electric displacement field arises naturally in the solution. This
electric displacement field is a consequence of the artificial change
of topology imposed on the system, in which the crystal blocks the
passage of ions, resulting in a uniformly polarized electrolyte. Our
analysis reveals that this is a consequence of the slab geometry
itself, rather than the lattice summation techniques that are
typically used to treat electrostatic interactions under periodic
boundary conditions. We argue that tinfoil Ewald approaches, which
impose zero electropotential difference in the electrolyte either side
of the crystal, are appropriate when attempting to model the polar
surfaces of crystals in solution, if the crystal is genuinely thin. In
contrast, if one is interested in crystals that are macroscopic in
extent, the electric displacement field can be chosen to enforce
$\sigma^{(\infty,\rm eq)}$, as demonstrated in
Ref.~\onlinecite{sayer2020macroscopic}.

Finally, our theoretical prediction (Eq.~\ref{eqn:fluct}) for the
fluctuations in the adsorbed surface charge density is intriguing, and
suggests that dynamics at the interface may change significantly as
the thickness of the crystal increases. Static energy calculations
already reveal that the behavior of thin films in contact with vacuum
can be complex, with structural relaxations that we have not
considered in this work permitting uncompensated polarity to a certain
extent \cite{goniakowski2007prediction}. The morphology of crystals is
generally determined by the relative growth rates along different
crystallographic directions
\cite{dandekar2013engineering}. Considering the potential interplay
between structural relaxations, and the thickness dependence of both
the equilibrium adsorbed surface charge density and its fluctuations,
one can easily imagine that the growth mechanisms of a polar surface
from either a melt or supersaturated solution will be a complex
affair. In addition, we have implicitly assumed $\sqrt{\mathcal{A}}\gg
nR$ throughout, where $\mathcal{A}$ is the surface area of the exposed
polar facets in the crystal (Fig.~\ref{fig:schematic1}a), whereas the
polar catastrophe has nontrivial dependence on crystal morphology
\cite{nosker1970polar,noguera2013polarity}. Future work will focus on
disentangling these different contributions to the growth mechanisms
of polar crystal surfaces.

\section{Methods}
\label{sec:Methods}

The simulation methodology closely resembles that of
Ref.~\onlinecite{sayer2020macroscopic}, and indeed, for the $R_{\rm
  c}=R$ system, we reused trajectories for $n=3$, $5$, $7$, $11$
and~$23$. All simulations used the SPC/E water model
\cite{BerendsenStraatsma1987sjc}, whose geometry was constrained using
the \texttt{RATTLE} algorithm \cite{andersen1983rattle} and the
Joung-Cheatham NaCl force field
\cite{joung2008determination}. Dynamics were propagated using the
velocity Verlet algorithm with a time step of 2\,fs. The temperature
was maintained at 298\,K with a Nos\'{e}-Hoover chain
\cite{shinoda2004rapid,tuckerman2006liouville}, with a 0.2\,ps damping
constant. The particle-particle particle-mesh Ewald method was used to
account for long-ranged interactions \cite{HockneyEastwood1988sjc},
with parameters chosen such that the root mean square error in the
forces were a factor $10^{5}$ smaller than the force between two unit
charges separated by a distance of 1\,\AA\cite{kolafa1992cutoff}. A
cutoff of 10\,\AA{} was used for non-electrostatic interactions.

The electrolyte comprised 600 water molecules and 20 NaCl ion
pairs. The crystal consisted of alternating layers of \ce{Na+} and
\ce{Cl-} ions, separated by $R=1.628$\,\AA, and each layer comprised
16 ions. The lateral dimensions of the simulation cell were
$L_x=15.952$\,\AA{} and $L_y=13.815$\,\AA{} along $x$ and $y$,
respectively. In the slab geometry with $n=3$ and $R_{\rm c}=R$, the
length of the simulation cell along $z$ was $L=94.841$\,\AA, and $L$
was increased with $n$ accordingly e.g. for $n=5$, $L$ was increased
by $2R$. Each simulation was 10\,ns long post equilibration. The
\texttt{LAMMPS} simulation package was used throughout
\cite{plimpton1995sjc}.


\begin{acknowledgments}
  I am grateful to Rob Jack, Michiel Sprik and Tom Sayer for many
  enlightening discussions on this topic, and Rick Remsing for reading
  a draft of the manuscript. This work was supported by a Royal
  Society University Research Fellowship (URF\textbackslash
  R1\textbackslash 211144). For the purpose of open access, I have
  applied a Creative Commons Attribution (CC BY) licence to any Author
  Accepted Manuscript version arising.
\end{acknowledgments}

\section*{Author Declarations}
\subsection*{Conflict of Interest}
The author has no conflicts to disclose. 

\section*{Data Availability}

The data that support the findings of this study are openly available
in the data repository at \url{https://doi.org/10.17863/CAM.83994}.

\appendix

\section{Derivation of $\sigma^{(n,\rm eq)}$ under tinfoil Ewald periodic boundary conditions}
\label{Appendix}

In the main article, $\sigma^{(n,\rm eq)}$ was derived for a
nonperiodic slab geometry (Eqs.~\ref{eqn:SigmaEqGen}
and~\ref{eqn:SigmaEqa}). In this appendix, an expression for
$\sigma^{(n,\rm eq)}$ is derived by considering the average molecular
charge distribution under periodic boundary conditions. The derivation
strongly hints that noncapacitive contributions arise from
approximating $\sigma^{(n,\rm eq)}$ as being confined to a single
plane close to the crystal's surface (Fig.~\ref{fig:schematic1}). The
derivation largely resembles that of Ref.~\onlinecite{hu2021comment},
but differs in that we do not consider a smeared average molecular
charge distribution. Moreover, we take the present derivation as
evidence that tinfoil Ewald sums provide a reliable estimate of
$\sigma^{(n,\rm eq)}$, rather than ion adsorption resulting from a
spurious treatment of electrostatic interactions.

When the average molecular charge density $\rho_{\rm c}$ only varies
along $z$, the average electrostatic potential under tinfoil Ewald
periodic boundary conditions is
\cite{wirnsberger2016non,pan2017effect,cox2018interfacial}
\begin{equation}
  \label{eqn:EwaldPhi}
  V(z) = 4\pi\int_{\rm cell}\!\mrm{d}z^\prime\,\rho_{\rm c}(z^\prime)J(z-z^\prime),
\end{equation}
with
\begin{equation}
  \label{eqn:EwaldJ}
  J(z) = \mrm{const.} + \frac{z^2}{2L} - \frac{|z|}{2},
\end{equation}
where $L$ is the length of the simulation cell along $z$. For
simplicity, let us first consider the case $n=1$. In the simulations
discussed in this article, the crystal ions do not move; approximating
these as uniformly charged planes we have,
\begin{equation}
  \rho_{\rm c}(z) = \rho_{\rm c}^{(\rm soln)}(z) + \sigma_0\big[\delta(z-R/2)-\delta(z+R/2)\big],
\end{equation}
where $\rho_{\rm c}^{\rm (soln)}$ is the average molecular charge
distribution arising from the electrolyte solution (both solvent and
ions). Let us assume that the crystal is centered at $z=0$. In the
region $R/2 < z \le L/2$ we then have
\begin{equation}
  V(z) = 4\pi\int_{\rm cell}\!\mrm{d}z^\prime\,\rho^{\rm (soln)}_{\rm c}(z^\prime)J(z-z^\prime) + 4\pi\sigma_0\bigg[\frac{-Rz}{L} + \frac{R}{2}\bigg],
\end{equation}
and an average electric field
\begin{equation}
  \hat{\mbf{z}}\cdot\mbf{E}(z) = -4\pi\int_{\rm cell}\!\mrm{d}z^\prime\,\rho^{\rm (soln)}_{\rm c}(z^\prime)\frac{\mrm{d}J(z-z^\prime)}{\mrm{d}z} + 4\pi\sigma_0\frac{R}{L}.
\end{equation}
At equilibrium, the electric field in the electrolyte vanishes, thus
\begin{equation}
  \label{eqn:EwaldRhoEqGen}
  \int_{\rm cell}\!\mrm{d}z^\prime\,\rho^{\rm (soln, eq)}_{\rm c}(z^\prime)\frac{\mrm{d}J(z-z^\prime)}{\mrm{d}z} = \sigma_0\frac{R}{L},
\end{equation}
where $z$ is understood to represent a point in the bulk
electrolyte. The equilibrium electrolyte charge distribution satisfies
Eq.~\ref{eqn:EwaldRhoEqGen}, no matter how complex its behavior. For a
general $\rho_{\rm c}^{(\rm soln, eq)}$, however,
Eq.~\ref{eqn:EwaldRhoEqGen} is not straightforward to analyze.

To simplify matters, we introduce a model for $\rho_{\rm c}^{(\rm
  soln)}$ whereby the adsorbed surface charge at each surface is
confined to a single plane:
\begin{equation}
  \rho_{\rm c}^{(\rm soln)}(z) \approx \sigma^{(n=1)}\big[\delta\big(z+(R+2\bar{\ell}\big)/2) - \delta\big(z-(R+2\bar{\ell})/2\big)\big].
\end{equation}
The relationship between $\bar{\ell}$ and the true average molecular
charge distribution is left unspecified, and it should not necessarily
be considered a physically meaningful length scale. For this model,
the average electrostatic potential for $(R+2\bar{\ell})/2<z\le L/2$
is
\begin{align}
  V(z) = &-4\pi\sigma^{(n=1)}\bigg[\frac{-(R+2\bar{\ell})z}{L} + \frac{R+2\bar{\ell}}{2}\bigg] \nonumber \\
  &+ 4\pi\sigma_0\bigg[\frac{-Rz}{L} + \frac{R}{2}\bigg],
\end{align}
while the electric field is
\begin{equation}
  \hat{\mbf{z}}\cdot\mbf{E}_{\rm e} =
  -4\pi\sigma^{(n=1)}\frac{R+2\bar{\ell}}{L} + 4\pi\sigma_0\frac{R}{L}.
\end{equation}
Again enforcing the equilibrium condition,
$\hat{\mbf{z}}\cdot\mbf{E}_{\rm e} = 0$, we find
\begin{equation}
  \label{eqn:EwaldSigmaEq1}
  \sigma^{(n=1,\rm eq)} = \frac{\sigma_0}{1+2\bar{\ell}/R},
\end{equation}
which agrees with Eq.~\ref{eqn:SigmaEqa}, provided that $\bar{\ell} =
\ell+a_{\rm nc}$. Comparing Eqs.~\ref{eqn:SigmaEqa},
\ref{eqn:EwaldRhoEqGen} and~\ref{eqn:EwaldSigmaEq1} suggests that
noncapacitive contributions largely account for the approximation of
treating the entire adsorbed surface charge density as if it were
confined to a single plane.

Following the same procedure for general $n$, we find
\begin{align}
  &\hat{\mbf{z}}\cdot\mbf{E}_{\rm e} = -4\pi\sigma^{(n)}\frac{(nR+2\bar{\ell})}{L} \nonumber \\ &- 4\pi\sigma_0\frac{R}{L}(-1)^{(n+1)/2}\sum_{j=1}^{(n+1)/2}(2j-1)(-1)^{j+1} \nonumber \\
  &= -4\pi\sigma^{(n)}\frac{(nR+2\bar{\ell})}{L} + 2\pi\sigma_0\frac{(n+1)R}{L}(-1)^{(n+1)/2}i^{n+1} \nonumber \\
  &= -4\pi\sigma^{(n)}\frac{(nR+2\bar{\ell})}{L} + 2\pi\sigma_0\frac{(n+1)R}{L}. 
\end{align}
To arrive at this result, we recall that $n+1$ is an even integer,
such that $i^{n+1}=(-1)^{(n+1)/2}$. At equilibrium,
\begin{equation}
  \sigma^{(n,\rm eq)} = \frac{(n+1)\sigma_0}{2nR + 4\bar{\ell}}.
\end{equation}
Again, we find agreement with Eq.~\ref{eqn:SigmaEqa}, provided that
$\bar{\ell} = \ell+a_{\rm nc}$. As we derived $\sigma^{(n,\rm eq)}$ by
enforcing $\hat{\mbf{z}}\cdot\mbf{E}_{\rm e} = 0$, and as $V(-L/2) =
V(L/2)$ (Eqs.~\ref{eqn:EwaldPhi} and~\ref{eqn:EwaldJ}), it immediately
follows that $V_{\rm a} = V_{\rm b}$ in the bulk electrolyte either
side of the slab (Fig.~\ref{fig:schematic2}, see also Fig.~4 of
Ref~\onlinecite{sayer2020macroscopic}). Note that $\sigma^{(n,\rm
  eq)}$ is independent of $L$. In the limit $L\to\infty$, any value of
$\sigma^{(n)}$ satisfies $\hat{\mbf{z}}\cdot\mbf{E}_{\rm e} = 0$,
which reflects the fact that the electric field external to a set of
infinite parallel plate capacitors vanishes. For
$\sigma^{(n)}\neq\sigma^{(n,\rm eq)}$, however, $V_{\rm a}\neq V_{\rm
  b}$, and the system becomes increasingly unstable as $n$ increases
(Eq.~\ref{eqn:fluct}).

\bibliography{./urf}


\end{document}